\documentclass[fleqn,usenatbib]{mnras}

\usepackage{newtxtext,newtxmath}

\usepackage[T1, T2A]{fontenc}
\usepackage{ae,aecompl}

\usepackage{graphicx}	% Including figure files
\usepackage{amsmath}	% Advanced maths commands
\usepackage{amssymb}	% Extra maths symbols

\usepackage{multirow}
\usepackage{adjustbox}

\usepackage[verbose]{placeins}
\usepackage{blindtext}

\title[The most luminous, un-obscured star-forming galaxy]{The discovery of the most UV-Ly$\alpha$ luminous star-forming galaxy: a young, dust- and metal-poor starburst with QSO-like luminosities}

\author[R. Marques-Chaves et al.]{
R. Marques-Chaves,$^{1,2,3}$ %\thanks{}
J. \'{A}lvarez-M\'{a}rquez,$^{1}$
L. Colina,$^{1,4}$
I. P\'{e}rez-Fournon,$^{2,3}$
\newauthor
D. Schaerer,$^{5,6}$
C. Dalla Vecchia,$^{2,3}$
T. Hashimoto,$^{7}$
C. Jim\'{e}nez-\'{A}ngel,$^{2,3}$ and
Y. Shu$^{8}$
\\
$^{1}$Centro de Astrobiolog\'ia (CSIC-INTA), Carretera de Ajalvir, 28850 Torrej\'on de Ardoz, Madrid, Spain; E-mail: rmarques@cab.inta-csic.es\\
$^{2}$Instituto de Astrof\'\i sica de Canarias, C/V\'\i a L\'actea, s/n, E-38205 San Crist\'obal de La Laguna, Tenerife, Spain\\
$^{3}$Universidad de La Laguna, Dpto. Astrof\'\i sica, E-38206 San Crist\'obal de La Laguna, Tenerife, Spain\\
$^{4}$International Associate, Cosmic Dawn Center (DAWN) \\
$^{5}$Geneva Observatory, University of Geneva, ch. des Maillettes 51, CH-1290 Versoix, Switzerland\\
$^{6}$CNRS, IRAP, 14 Avenue E. Belin, 31400 Toulouse, France\\
$^{7}$Tomonaga Center for the History of the Universe (TCHoU), Faculty of Pure and Applied Sciences, University of Tsukuba, Tsukuba, Ibaraki 305-8571, Japan\\
$^{8}$Institute of Astronomy, University of Cambridge, Madingley Road, Cambridge CB3 0HA, UK
}

\date{}

\begin{document}
\label{firstpage}
\pagerange{\pageref{firstpage}--\pageref{lastpage}}
\maketitle

% Abstract of the paper
\begin{abstract}
We report the discovery of BOSS-EUVLG1 at $z=2.469$, by far the most luminous, almost un-obscured star-forming galaxy known at any redshift. First classified as a QSO within the Baryon Oscillation Spectroscopic Survey, follow-up observations with the Gran Telescopio Canarias reveal that its large luminosity, $M_{\rm UV}\simeq-$24.40 and log($L_{\rm Ly \alpha} / \rm erg~s^{-1})\simeq 44.0$, is due to an intense burst of star-formation, and not to an AGN or gravitational lensing. 
BOSS-EUVLG1 is a compact ($r_{\rm eff} \simeq$1.2~kpc), young (4-5~Myr) starburst with a stellar mass log($M_{*}/M_{\odot}$)=10.0$\pm$0.1 and a prodigious star formation rate of $\simeq$1000~$M_{\odot}$~yr$^{-1}$. However, it is metal- and dust-poor (12+log(O/H)=8.13$\pm$0.19, E(B-V)$\simeq$0.07, log($L_{\rm IR}$/$L_{\rm UV}) < -1.2$), indicating that we are witnessing the very early phase of an intense starburst that has had no time to enrich the ISM. 
BOSS-EUVLG1 might represent a short-lived ($<$100~Myrs), yet important phase of star-forming galaxies at high redshift that has been missed in previous surveys. 
Within a galaxy evolutionary scheme, BOSS-EUVLG1 could likely represent the very initial phases in the evolution of massive quiescent galaxies, even before the dusty star-forming phase.

\end{abstract}


%%%%%%%%%%%%%%%%%%%%%%%%%%%%%%%%%%%%%%%%%%%%%%%%%%

%%%%%%%%%%%%%%%%% BODY OF PAPER %%%%%%%%%%%%%%%%%%

\section{Introduction}

Over the past decades, deep extragalactic surveys have been designed to identify star-forming galaxies (SFGs) at high redshift ($z>2$), such as Lyman break galaxies (LBGs) and Lyman-$\alpha$ emitters (LAEs), providing a wealth of information on their physical properties \citep[e.g.,][]{shapley2003} and investigating their space density as a function of luminosity through luminosity functions \citep[LFs; e.g.,][]{reddy2009, sobral2018a}.
The bright-end of LFs of the UV and Ly$\alpha$ emission probes massive star formation, however, identifying such UV/Ly$\alpha$-luminous SFGs is challenging due to their very low space density. The most efficient SFGs consume their gas faster, implying short timescales of their luminous phases, thus low observable abundances. Furthermore, star formation produces dust, so that the most vigorous and intrinsically luminous SFGs are expected to have their emission at short wavelengths heavily obscured (e.g., dusty star-forming galaxies, DSFGs; \citealt{casey2014}).  
Probing the bright-end of LFs requires thus very large area surveys and, in addition, spectroscopic follow-up to distinguish SFGs from the much more abundant active galactic nuclei \citep[AGNs; e.g.,][]{sobral2018b}. %, spinoso2020}. 
However, the widest dedicated surveys at $z\simeq 2-3$ with a few deg$^{2}$ (e.g., \citealt{ouchi2008, sobral2018b}) have failed to discover SFGs more luminous than 2 times the typical luminosity ($L^{*}$) of UV and Ly$\alpha$ of LBGs/LAEs, implying maximal un-obscured luminosities by a starburst of $M_{\rm UV}$=$-$21.5 and log($L_{\rm Ly\alpha}/$erg~s$^{-1}$)=43.3 \citep[][]{sobral2018b}.

In this Letter we study SDSS J122040.72+084238.1, hereafter BOSS-EUVLG1, located at ($\alpha$,~$\delta )_{\rm J2000} = (185.1697^{\circ},8.7106^{\circ}$) with a redshift $z$=2.469. It was discovered as part of our search for Extremely UV-Luminous Galaxies (EUVLGs, $M_{\rm UV}$<$-$23) within the $\sim 9300$~deg$^{2}$-wide extended Baryon Oscillation Spectroscopic Survey \citep[eBOSS:][]{abolfathi2018} of the Sloan Digital Sky Survey \citep[SDSS:][]{eisenstein2011}. BOSS-EUVLG1 is 
the most luminous, un-obscured star-forming galaxy known in the Universe by far, with $M_{\rm UV}$=$-$24.4 and log($L_{\rm Ly\alpha}/$erg~s$^{-1}$)=44.0.  
Throughout this work, we assume a \cite{salpeter1955} initial mass function (IMF) and a concordance cosmology with $\Omega_{\rm m} = 0.274$, $\Omega_{\Lambda} = 0.726$, and $H_{0} = 70$ km s$^{-1}$ Mpc$^{-1}$. All magnitudes are given in the AB system.

\section{Discovery and Follow-up Observations}

In our survey ''BOSS Extremely UV-Luminous Galaxies'' (BOSS-EUVLG) we have identified a total of $\sim$70 SFGs at 2<$z$<4 with $-$24.5<$M_{\rm UV}$<$-$23, being BOSS-EUVLG1 the brightest and most luminous one, and the subject of this work. The sample and the selection techniques will be presented in a future work (Marques-Chaves et al. in prep.).
Its BOSS spectrum (plate-mjd-fiberid: 5397-55944-626) 
shows features characteristic of an un-obscured, luminous SFG without any evidence so far of a hidden AGN, showing narrow C~{\sc iii}] emission ($\sim$200~km~s$^{-1}$ full width half maximum, FWHM) and strong P-Cygni stellar profiles. The systemic redshift $z$=2.4694$\pm$0.0004 is measured from the C~{\sc iii}] doublet.

\begin{figure*}
  \centering
  \includegraphics[width=0.98\textwidth]{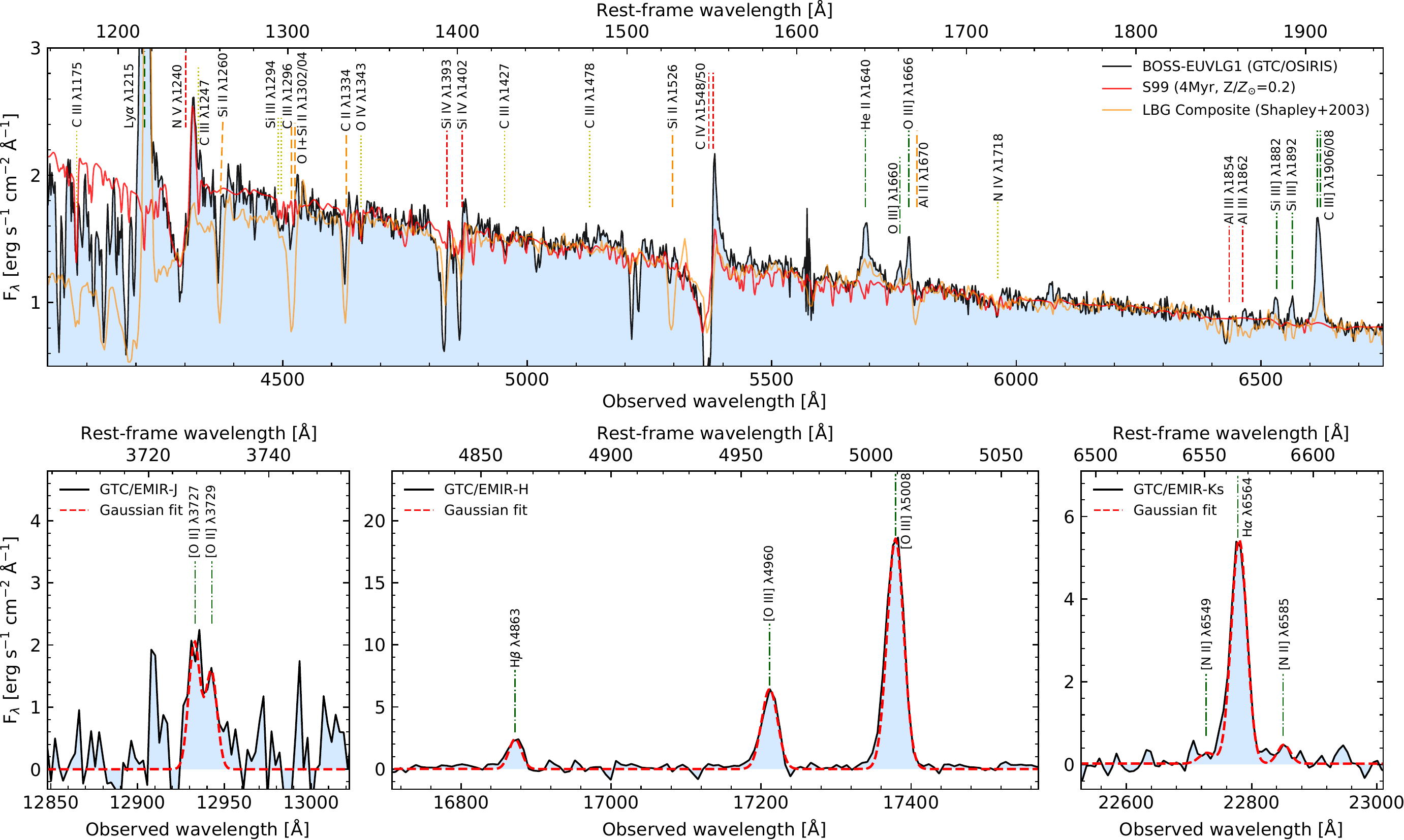}
  \caption{GTC rest-frame UV (up) and optical (down) spectra of BOSS-EUVLG1. Ticks mark the positions of main features associated with stellar P-Cygni wind lines (red) and photospheric absorption (yellow), nebular emission (green) and ISM absorption (orange). The best-fit S99 model with age of 4~Myr and $Z/Z_{\odot}$=1/5 is shown in red (E(B-V)=0.07). The $z \sim 3$ LBG composite at the redshift of BOSS-EUVLG1 is shown in orange \citep{shapley2003}.}
  \label{fig1}
\end{figure*}

Optical and near-IR long-slit spectroscopy were obtained with the Gran Telescopio Canarias (GTC) OSIRIS and EMIR instruments, respectively, as part of the programs GTC67-17A and GTCMULTIPLE2F-18A (PI: R.~Marques-Chaves).
OSIRIS observations were performed using the R1000B grism ($\rm R\sim 700$) with a total integration time of 50~min. The data were reduced with standard {\sc Iraf} tasks. Near-IR spectra were obtained using the $J$ ($\rm R\sim 3000$) and $HK$ grisms ($\rm R\sim 700$) with a total observing time of 27 and 43~min, respectively. Reduction was performed using the EMIR pipeline. Optical and near-IR spectra were flux calibrated using standards stars, and their fluxes matched those obtained from photometry. 
Optical and near-IR spectra of BOSS-EUVLG1 are shown in Fig. \ref{fig1}.

Broad-band imaging were also obtained in the near-IR with EMIR, using $J$, $H$, and $K_{\rm s}$ filters. These were flux calibrated against 2MASS stars in the field. We also used archival Canada France-Hawaii Telescope (CFHT) optical images in $U$ to $Z$ bands processed with the MegaPipe image staking pipeline \citep{gwyn2008}. Both optical and near-IR images have sub-arsec seeing conditions $\simeq 0.55^{\prime \prime}-0.9^{\prime \prime}$ FWHM). 
We use aperture photometry with a diameter of 2.5$\times$FWHM. BOSS-EUVLG1 is also detected in the mid-IR in the sixth year NEOWISE 2020 data release \citep[][]{meisner2017} at $3.4\mu$m ($W1$=20.60$\pm$0.11), but not at $4.6\mu$m ($3\sigma$ limit $W2$>21.20).
Finally, we observed the dust continuum toward BOSS-EUVLG1 at 1.2~mm using the Atacama Large submillimeter/Millimeter Array (ALMA: project ID: 2018.1.00932.S; PI: R.~Marques-Chaves) with a total on source time of 5.6~min.
Data reduction was performed using version 5.4.0 of the Common Astronomy Software Applications ({\sc casa}) package. BOSS-EUVLG1 is not detected with a $3\sigma$ limit of 37$\mu$Jy (beam size of $0.62^{\prime \prime} \times 0.61^{\prime \prime}$).

\section{Results and Discussion}

\subsection{The Nature of BOSS-EUVLG1}

The most striking property of BOSS-EUVLG1 is its brightness, and corresponding large luminosities at $z\simeq2.469$, in both the rest-frame UV continuum ($R \simeq 20.8$) and nebular emission (fluxes in Ly$\alpha$ and H$\alpha$ above $10^{-15}$~erg~s$^{-1}$~cm$^{-2}$, Table \ref{table1}). Such apparent fluxes are much larger than those observed in typical SFGs (factors $\approx 30$; e.g., \citealt{reddy2009, sobral2018a}) and are only comparable to those observed in highly magnified lensed systems \citep[e.g.,][]{shu2016b, marques2020} or in QSOs \citep[e.g.,][]{paris2014}.

\begin{table*}
\begin{center}
\caption{Integrated fluxes and rest-frame equivalent widths (units of $10^{-17}$~erg~s$^{-1}$~cm$^{-2}$ and \AA, respectively) of main nebular emission lines in BOSS-EUVLG1. The values of He~{\sc ii}~1640\AA{ }refer to the narrow component. \label{table1}}
\begin{tabular}{l c c c c c c c c c c }
\hline \hline
\smallskip
\smallskip
Line  &  Ly$\alpha$ &  He~{\sc ii} & C~{\sc iii}] & [O~{\sc ii}] & H$\beta$ &  [O~{\sc iii}]  & [O~{\sc iii}] & H$\alpha$ & [N~{\sc ii}] \\
$\lambda_{\rm rest}^{\rm vac}$ (\AA) & 1215.57 & 1640.42 & 1906.68,1908.68 & 3727.09,3729.88 & 4862.68 & 4960.30 & 5008.24 & 6564.61 & 6585.27 \\
\hline 
Flux & $194 \pm 17$ & $4.6 \pm 0.4$ & $17.3 \pm 0.9$ & $18\pm4$, $13\pm3$ & $61 \pm 5$ & $176\pm6$ & $562 \pm 7$ & $190 \pm 5$ & $<12 (3\sigma)$\\
$EW_{0}$ & $ 22 \pm 3$ & $1.0 \pm 0.1$ & $4.4 \pm 0.2$ & --- & $115 \pm 25$ & $350 \pm 70$ & $1125 \pm 225$ & $685 \pm 185$ & $<43 (3\sigma)$ \\
\hline 
\end{tabular}
\end{center}
\end{table*}

\subsubsection{Neither lensed galaxy nor luminous AGN}

\begin{figure*}
  \centering
  \includegraphics[width=0.98\textwidth]{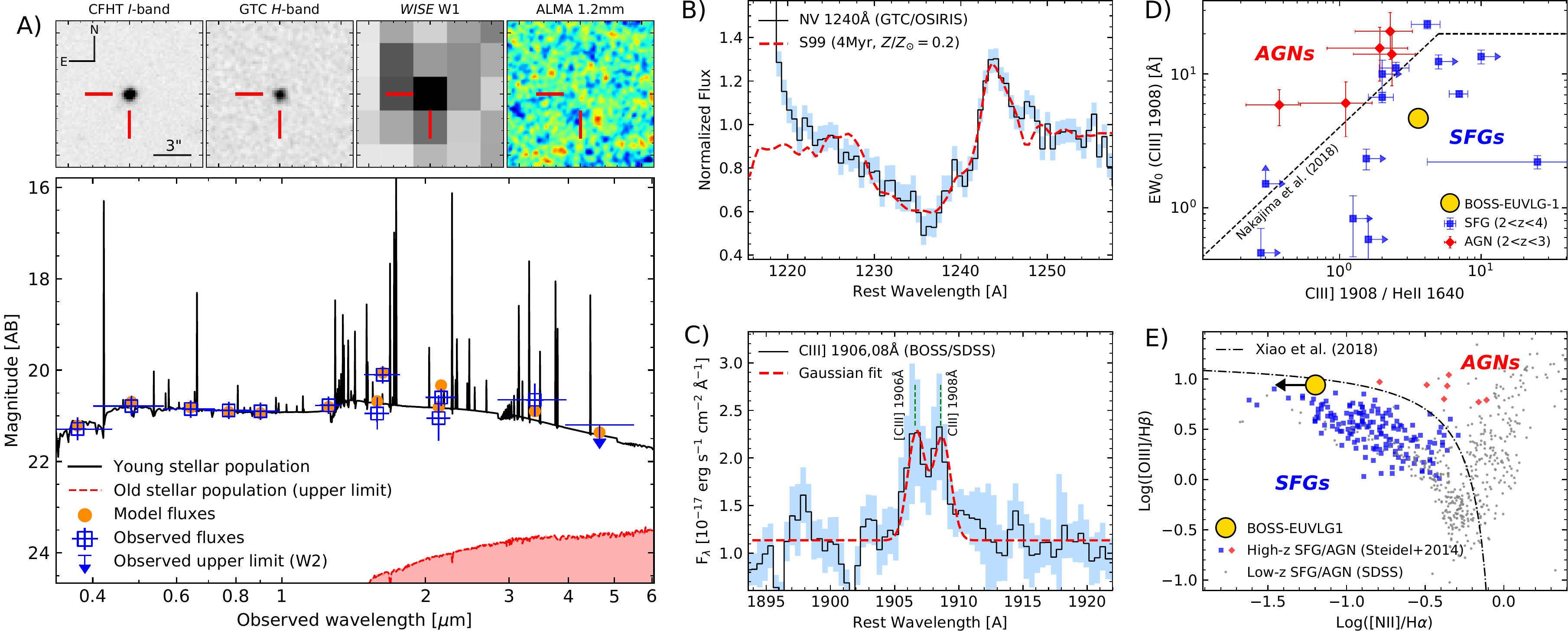}
  \caption{Summary of the main results. A: cutout images of BOSS-EUVLG1 (12$^{\prime \prime}$x12$^{\prime \prime}$) in selected bands from optical to mm (top) and best-fit SED of BOSS-EUVLG1, characterized by a young and intense burst of star formation (down); B) comparison of the normalized GTC spectrum (black) with the best-fit S99 model (4~Myr and $Z/Z_{\odot}$=1/5, dashed red) around the wind line N~{\sc v}; C: [C~{\sc iii}] nebular emission as seen in the BOSS spectrum (black). The resolved lines have widths of $\simeq$200~km~s$^{-1}$ FWHM; D and E: BOSS-EUVLG1 (yellow circle) in the rest-frame UV and optical line diagnostics (\citealt{nakajima2018}, panel D; and \citealt{xiao2018}, BPT, panel E).}
  \label{fig2}
\end{figure*}

The lack of lensing structures, such as multiple images or arc-like morphologies, and foreground galaxies in optical and near-IR images (Fig. \ref{fig2} A) make unlikely that BOSS-EUVLG1 is being magnified by gravitational lensing. It shows a compact morphology that is only barely resolved in images taken with very good seeing conditions. Using the CFHT $I$-band (seeing $\simeq 0.55^{\prime \prime}$~FWHM) and {\sc Galfit} \citep[][]{peng2002}, the light distribution of BOSS-EUVLG1 is well modeled with a Sersic profile with an effective radius $r_{\rm eff} = 0.15^{\prime \prime} \pm 0.04^{\prime \prime}$ ($1.2 \pm 0.3$~kpc) and an index $n=1.3\pm0.4$. 

A strong contribution of an AGN is highly unlikely. The rest-frame UV and optical spectra (Fig. \ref{fig1}) show several spectral features that are common in SFGs. In particular, stellar wind P-Cygni profiles (N~{\sc v}~1240\AA{ } or C~{\sc iv}~1550\AA) and photospheric absorption lines (e.g., C~{\sc iii}~1427\AA{ }or N~{\sc iv}~1718\AA) are clearly detected with high significance, indicating that the UV luminosity is dominated by stellar emission (at least >70\%) rather than AGN emission (Fig. \ref{fig2} B). 
Nebular emission also presents intrinsic narrow spectral profiles, e.g.,  $\simeq$200~km~s$^{-1}$~FWHM for C~{\sc iii}] and [O~{\sc ii}]~3726,3729\AA{ }in the BOSS and EMIR/J spectra, respectively (Figs. \ref{fig1} and \ref{fig2} C).
The exception is the He~{\sc ii}~1640\AA{ }emission line, that shows two clearly distinguished components, one narrow from nebular emission and another one much broader ($\simeq 2600$~km~s$^{-1}$ FWHM) likely associated with the presence of a large number of Wolf-Rayet stars \citep[][]{schaerer1998}. 
The mid-IR color $W1-W2$<$-$0.60 also disfavours the presence of a strong AGN component, which typically shows $W1-W2$>0.2 at $z\approx 2.5$ from the dust torus \citep[][]{assef2013}.
We also use rest-frame UV and optical line diagnostics to investigate the contribution of an AGN to the nebular emission (Fig. \ref{fig2} D and E). Following the star-formation and AGN models of \cite{nakajima2018}, the observed line ratio of C~{\sc iii}]/He~{\sc ii}$ = 3.6 \pm 0.4$ and the C~{\sc iii}] rest-frame equivalent width 4.4$\pm$0.2\AA{ }(Table \ref{table1}) place BOSS-EUVLG1 in the star-formation region. The line ratios of log([O~{\sc iii}]/H$\beta$)=0.96$\pm$0.03 and log([N~{\sc ii}]/H$\alpha$)<$-$1.20 (3$\sigma$) place it far away from the locus of low-$z$ SFGs, but close to that expected for high-$z$ SFGs \citep[e.g.,][]{steidel2014} with high ionizing parameters, $U$, and large electron density, $n_{\rm e}$ \citep[][]{xiao2018}. Indeed, the large [O~{\sc iii}]/[O~{\sc ii}]$\simeq$18 and [O~{\sc ii}]~3729/3726$\simeq$0.7 (Table \ref{table1}) suggest such extreme properties (log($U$)>$-$2 and log($n_{\rm e}$/cm$^{-3}$)>10$^{3}$, Marques-Chaves in prep.).

\subsubsection{BOSS-EUVLG1 as an extremely luminous starburst}

Our results indicate that the large luminosity is not arising from an AGN and that the effect of gravitational lensing is highly unlikely. Instead, the large luminosities arise from a large number of OB stars present in the first stages of an intense burst of star-formation. 
We compare the wind line profiles of N~{\sc v} and C~{\sc iv} to those obtained with the spectral synthesis code {\sc Starburst99} \citep[S99:][]{leitherer1999} to infer the age and metallicity of the stellar population \citep[details in][]{marques2018, marques2020}.  
P-Cygni profiles are well modeled by a $Z/Z_{\odot} \simeq 1/5$ and $ \simeq $4~Myr burst of star formation (Fig. \ref{fig1}). We infer a color excess of the stellar continuum $E (B-V)_{*}$=0.07$\pm$0.02 (\citealt{calzetti2000}) by comparing the observed UV slope $\beta^{\rm obs}_{\rm UV}$=$-$2.16$\pm$0.05 with that from the best-fit S99 model ($\beta^{\rm syn}_{\rm UV}$=$-$2.44).

BOSS-EUVLG1 shows a flat spectral energy distribution (SED) from optical to the mid-IR ($R-W2$<$-$0.35), except in the near-IR $H$ and $K_{\rm s}$ bands where a flux excess is clear identified (e.g., $J-H$=0.67$\pm$0.13, Fig. \ref{fig2} A) due to the strong contribution of rest-frame optical H$\beta$+[O~{\sc iii}] and H$\alpha$ emission to the $H$ and $K_{\rm s}$ bands.
We perform SED-fitting with CIGALE \citep{Boquien2019} using the broad-band photometry (from $U$ to W2 bands), two synthetic bands generated from the collapsed EMIR spectra that sample the rest-frame optical continuum free from nebular emission, and the observed H$\beta$+[O~{\sc iii}] and H$\alpha$ line fluxes. The star-formation history (SFH) is modeled using two components; (i) instantaneous SFH with ages ranging from 1-10~Myr, and (ii) a exponentially declining SFH with ages $>1$~Gyr. We adopted the stellar population models from \cite{bruzual2003}, and the \cite{calzetti2000} dust attenuation law. The metallicity is fixed to $Z/Z_{\odot}$=1/5. 
The best-fit model, shown in Fig. \ref{fig2} (A), indicates that BOSS-EUVLG1 is dominated by a young stellar burst characterized by a SFR=1060$\pm$100~$M_{\odot}$~yr$^{-1}$, an age of 5$\pm 1$~Myr, and a stellar mass of $\log (M_{*}/M_{\odot}$)=10.0$\pm$0.1. This yields to a high specific SFR (sSFR=SFR/$M_{*}$) of 106$\pm$15~Gyr$^{-1}$. 
The old stellar population is not well constrained, yet it should be faint as the SED from $U$ to W2 ($0.10-1.33\mu$m rest-frame) is dominated by the emission from the young burst, and should be less massive than log($M_{*}^{\rm old}/M_{\odot}$)<10.2 otherwise a >3$\sigma$ detection at 4.6$\mu$m would be expected (Fig. \ref{fig2}~A).

Turning to the nebular emission, BOSS-EUVLG1 shows intense lines in the rest-frame UV and optical (Fig. \ref{fig1} and Table \ref{table1}). The detection of nebular He~{\sc ii} emission and the large [O~{\sc iii}]/[O~{\sc ii}]$\simeq$18 are indicative of a hard ionizing spectrum. The rest-frame equivalent widths $EW_{0}$(H$\beta$)=115$\pm$25\AA{ }and $EW_{0}$(H$\alpha$)=685$\pm$185\AA{ }are consistent with S99 burst models with age $\simeq 4$~Myrs.
Using the H$\alpha$ flux we measure a total $\rm SFR (H\alpha) = 955 \pm 135$~M$_{\odot}$~yr$^{-1}$ \citep{kennicutt1998}, after correction for dust obscuration using the observed line ratio H$\alpha$/H$\beta = 3.10 \pm 0.18$ ($E(B-V)_{\rm neb}=0.07\pm0.05$, \citealt{calzetti2000}). 
The $\rm SFR (H\alpha)$ is significantly larger than that measured using the UV luminosity, $\rm SFR (UV) = 680\pm125$~M$_{\odot}$~yr$^{-1}$ \citep{kennicutt1998}, but expected for very young starbursts like BOSS-EUVLG1, as H$\alpha$ and UV indicators trace different time-scales \citep[$\sim$10 and 100~Myrs, respectively; e.g.,][]{flores2020}.
The Ly$\alpha$ emission is not spatially resolved (<0.8$^{\prime \prime}$~FWHM). Its spectral profile shows two different velocity components. The main component ($\simeq$90\% of the total flux) appears redshifted relative to the rest velocity by 260$\pm$60~km~s$^{-1}$ and  shows an intrinsic FWHM of $\simeq$480~km~s$^{-1}$. A second and fainter component (10\% of the total flux) is found blueshifted by $\simeq -650$~km~s$^{-1}$. Assuming case-B recombination we measure a total Ly$\alpha$ escape fraction $f_{\rm esc} \rm (Ly\alpha)$=0.10$\pm$0.2, in line with the empirical relation derived by \cite{sobral2019} considering the $EW_{0} \rm (Ly\alpha)$=22$\pm$3~\AA.
We measure a dust-corrected line ratio $R$23=([O~{\sc ii}] + [O~{\sc iii}])/H$\beta=12.5\pm3.3$, yielding to a nebular metallicity 12+log(O/H)=8.13$\pm$0.19 \citep[][]{pilyugin2005}, compatible with that measured for the young stellar population.
The ISM of BOSS-EUVLG1 also appears mostly ionized, showing weak low-ionization ISM lines (e.g., $EW_{0}\simeq 0.66$\AA{ }for Si~{\sc ii}~1260\AA), but strong high-ionization ones (e.g., $\simeq$2.55\AA{ } for Si~{\sc iv}~1393\AA). These lines have their centroids blueshifted by $-$320$\pm$80~km~s$^{-1}$ relative to the systemic velocity, indicative of strong large scale outflows.

The large SFR of BOSS-EUVLG1 (from H$\alpha$ and SED) is only comparable to those measured in hyper-luminous DSFGs, and should be detected in the far-IR if large amounts of dust have been already formed. However, it is not detected in dust continuum in ALMA with a $3\sigma$ limit of 37~$\mu$Jy at 1.2mm (Band 6). Using the far-IR SED of ALESS galaxies ($T_{\rm dust} \simeq 43 K$; \citealt{dacunha2015}), this limit yields to an integrated far-IR luminosity log($L_{\rm IR} / L_{\odot}) < 10.91$ (from $8-1000 \mu$m) and a dust mass $M_{\rm dust} / M_{\odot} < 1.9\times 10^{7}$ ($3\sigma$), implying a log($L_{\rm IR}$/$L_{\rm UV}) < -1.2$.
While a large $T_{\rm dust}$ would lead to larger upper limits on $L_{\rm IR}$ and $M_{\rm dust}$, a low dust content is actually expected if a very young age of the burst is considered. Assuming a constant $\rm SFR = 955 \pm 135$~M$_{\odot}$~yr$^{-1}$ over 10~Myrs, we would expect a $M_{\rm dust} / M_{\odot} = (2.0 \pm 1.0) \times 10^{7}$ formed by supernovae (\citealt{gall2018}, assuming no dust destruction), still compatible with the non-detection of ALMA dust continuum. The non-detection of dust also suggests minor past starburst episodes, and that the bulk of stellar mass was formed in just few Myrs, in line with the results from SED-fitting.

\subsection{BOSS-EUVLG1: the prototype of the first and intense bursts of star-formation at high redshift}

\begin{figure}
  \centering
  \includegraphics[width=0.48\textwidth]{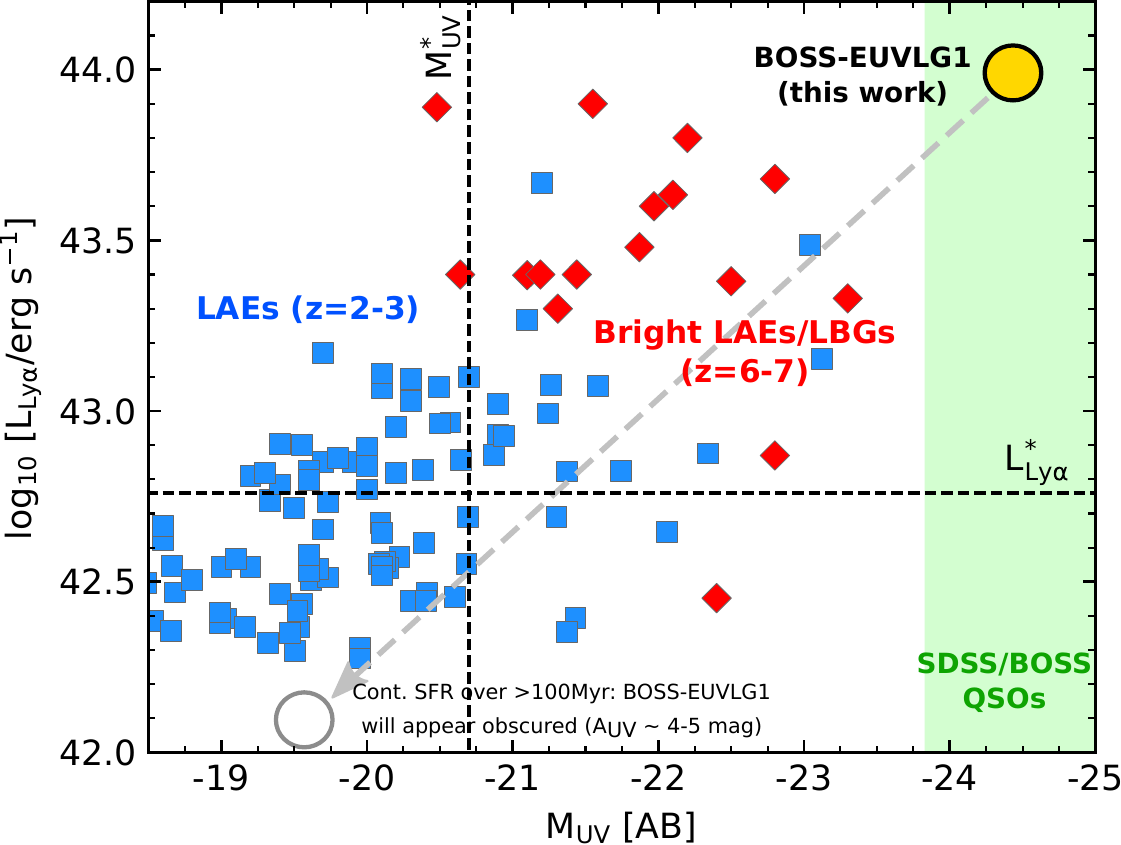}
  \caption{Ly$\alpha$ luminosity and UV absolute magnitude of BOSS-EUVLG1 (yellow) and other LAEs/LBGs at similar redshift \citep[blue;][]{ouchi2008, sobral2018b, marques2017, marques2020} and at $z\sim 6-7$ \citep[red;][]{sobral2015, matthee2017, matsuoka2018, shibuya2018, hashimoto2019}. Horizontal and vertical dashed lines represent the typical Ly$\alpha$ and UV luminosities of $z\sim 2-3$ LAEs and LBGs \citep[][]{reddy2009, sobral2018a}. If the SFR of BOSS-EUVLG1 is prolonged over $\gtrapprox$100~Myr, its UV and Ly$\alpha$ luminosity will likely become heavily obscured (empty circle).
  }
  \label{fig3}
\end{figure}

BOSS-EUVLG1 is by far the most luminous, dust-poor star-forming galaxy discovered at any redshift, with $M_{\rm UV} = -24.40\pm0.05$ and log($L_{\rm Ly \alpha} / \rm erg~s^{-1})=44.0\pm0.1$. 
Fig. \ref{fig3} compares the Ly$\alpha$ and UV luminosities of BOSS-EUVLG1 with those from other LAEs and LBGs, including the most luminous ones known at similar \citep[][]{marques2017, marques2020} and higher redshifts \citep[][]{sobral2015, matsuoka2018, shibuya2018}.
Our findings show that star-formation alone can produce almost un-obscured luminosities comparable to QSOs, well above (by factors of $\simeq$15) the proposed limit for star-formation UV/Ly$\alpha$ luminosities at $z=2-3$ \citep[][]{sobral2018b}.

The properties of BOSS-EUVLG1 are still intriguing and different from those observed in other galaxies. Its large luminosity and SFR ($\simeq$1000~$M_{\odot}$yr$^{-1}$), yet residual dust attenuation (E(B-V)$\simeq$0.07) and low metallicity ($Z/Z_{\odot}\simeq 1/5$), are clearly in contrast with those found in other samples, where the most luminous and vigorous SFGs are statistically more dusty and metal-enriched \citep[e.g.,][]{bouwens2009, casey2014}. A natural explanation, supported by our results, is that we are witnessing the very early phase ($\lesssim$10~Myrs) of the formation of a massive galaxy that has had no time to enrich its ISM with metals and dust. 

This phase should be, however, short-lived. If the star-formation is quenched, BOSS-EUVLG1 will appear redder and UV-faint in some tens Myrs ($\sim$3~mag fainter in 30~Myrs, according to S99), as the most luminous OB stars will no longer be present. It could thus represent the progenitors of compact ellipticals found at $z\sim 2$ ("red nuggets"; e.g., \citealt{dekel2014}). 
On the other hand, if it lies in a gas-rich environment and the SFR is prolonged over a few hundreds Myrs, large amount of dust should be produced ($M_{\rm dust}/M_{\odot} \sim (2 -4) \times 10^{8}$ formed by supernovae \citealt{gall2018}), approaching those observed in ultra- and hyper-luminous DSFGs ($M_{\rm dust} /M_{\odot} \sim 10^{8}-10^{9}$; e.g., \citealt{santini2010}). In such case, BOSS-EUVLG1 could represent the very early phase ($\sim 10$~Myr) of DSFGs, becoming a very bright source in the far-IR, but heavily obscured in the UV in just $\sim$100~Myrs ($A_{\rm UV}\simeq$ 4-5mag, Fig. \ref{fig3}; e.g., \citealt{dacunha2015}). Such rapid change from UV-bright to IR-bright is supported by simulations \citep[][]{arata2019}.

The gas content and the efficiency to keep the vigorous SFR over a prolonged period will definitively dictate different fates for BOSS-EUVLG1, but in any case its extremely, dust-poor luminous phase is likely short, <100~Myrs. 
Moreover, such phase should be only visible during the first, major episodes of star-formation, otherwise the dust produced in past SFHs would likely obscure the UV and nebular radiation. This implies a very low observable abundance of these luminous sources and might explain why such galaxies have not been discovered before. 
The sharp statistical transition between SFGs and AGNs at $z \sim 2-3$ around $M_{\rm UV}$=$-$21.5 and log($L_{\rm Ly\alpha}/$erg~s$^{-1}$)=43.3 found by \cite{sobral2018b} already suggests that extremely UV and Ly$\alpha$ luminous sources like BOSS-EUVLG1 should be very rare as AGNs are the norm at these luminosities. Considering the other $\sim$70 EUVLGs at $z\sim 2-3$ identified in the $\sim$9300~deg$^{2}$-wide eBOSS survey, we estimate a number density of $\sim$10$^{-9}$~Mpc$^{-3}$, three orders of magnitude lower than that measured in QSOs at similar redshifts and $M_{\rm UV}$ (e.g., \citealt{palanque2016}).
This, in turn, does not necessarily mean that luminous phases are intrinsically rare. SFGs at $z$>6 are thought to have bursty SFHs, due to the high merger rate and accretion of primeval gas at such early times, thus expected to experience similar luminous phases in their first episodes of star-formation.
In fact, some exceptional luminous SFGs (both in the UV and Ly$\alpha$), have already been discovered at $z$>6 (e.g., \citealt{sobral2015, matsuoka2018, hashimoto2019}, Fig. \ref{fig3}).  
Similar to BOSS-EUVLG1, their SEDs, intense nebular emission, high sSFRs, and large, still un-obscured UV luminosities highly suggest that they are undergoing their first, major bursts of star-formation.

\section{Summary and Final remarks}

This Letter reports the discovery of BOSS-EUVLG1 at $z=2.4694\pm0.0004$, the most luminous, dust-poor star-forming galaxy known at any redshift by far. 
BOSS-EUVLG1 is a compact ($r_{\rm eff}\simeq$1.2~kpc), very young (4-5~Myr) starburst with log($M_{*}/M_{\odot}$)=10$\pm$0.1 and a prodigious  SFR=955-1060~$M_{\odot}$~yr$^{-1}$ (from H$\alpha$ and SED-fitting, respectively) and sSFR$\simeq$100~Gyr$^{-1}$. 
It exhibits QSO-like luminosities, $M_{\rm UV} \simeq -24.40$ and log($L_{\rm Ly \alpha} / \rm erg~s^{-1})\simeq 44.0$, likely arising from a large number of massive stars present in the first stages of the intense starburst. Its SED is dominated by the young starburst without a relevant old stellar component, yielding to intense nebular emission in the rest-frame UV (e.g., Ly$\alpha$, He~{\sc ii} or C~{\sc iii}]) and optical ($EW_{0}$(H$\beta$+[OIII])$\simeq$1600\AA{ } and $EW_{0}$(H$\alpha$)$\simeq$680\AA). Despite the large luminosity and SFR, BOSS-EUVLG1 is metal-poor (12+log(O/H)=8.13$\pm$0.19) and almost un-obscured (E(B-V)$\simeq$0.07) with no detection of dust continuum (log($L_{\rm IR}$/$L_{\rm UV}) < -1.2$), indicating that we are witnessing the very early phase of an intense starburst that has had no time to enrich the ISM with metals and dust. BOSS-EUVLG1 could be identified as a prototype of the first and intense episodes of star-formation, likely representing a short, initial phase in the evolution of compact ellipticals or DSFGs found at $z\sim 2$. Such phase would imply a fast mode -- almost instantaneous at cosmic scales -- of the stellar mass build-up and chemical and dust enrichment histories of SFGs in the early Universe. In future works, we will investigate in more detail the physical condition and properties of BOSS-EUVLG1, as well as  present a large sample of other $\sim$70 EUVLGs discovered within the  eBOSS survey.

\section*{Acknowledgements}

The authors thank the anonymous referee for useful comments. 
Based on observations made with the Gran Telescopio Canarias (GTC)  installed in the Spanish Observatorio del Roque de los Muchachos of the Instituto de Astrof\'{i}sica de Canarias, in the island of La Palma. This paper makes use of the following ALMA data: ADS/JAO.ALMA\#2018.1.00932.S. ALMA is a partnership of ESO (representing its member states), NSF (USA) and NINS (Japan), together with NRC (Canada), MOST and ASIAA (Taiwan), and KASI (Republic of Korea), in cooperation with the Republic of Chile. The Joint ALMA Observatory is operated by ESO, AUI/NRAO and NAOJ. 
R.M.C., J.A.M., L.C., I.P.F., C.D.V. and C.J.A acknowledge support from the Spanish State Research Agency (AEI) under grant numbers ESP2015-65597-C4-4-R, ESP2017-86852-C4-2-R, RyC-2015-18078,  PGC2018-094975-B-C22, and MDM-2017-0737 Unidad de Excelencia ''Mar\'{i}a de Maeztu''- Centro de Astrobiolog\'{i}a (CSIC-INTA).

\section*{Data availability}
The data underlying this article will be shared on reasonable request to the corresponding author.

\bibliographystyle{mnras}
\input{main.bbl}

\label{lastpage}
\end{document}